# Precipitation during high temperature aging of Al−Cu alloys: a multiscale analysis based on first principles calculations

H. Liu[a], I. Papadimitriou[b], F. X. Lin[c], J. LLorca[b,d,1]

[a] Department of Materials Engineering, KU Leuven, Kasteelpark Arenberg 44 B, Leuven, Belgium
[b] IMDEA Materials Institute, C/Eric Kandel 2, Getafe 28906 – Madrid, Spain
[c] Institute of Mechanics, Materials and Civil Engineering. Université catholique de Louvain, Avenue Georges Lemaître 4-6, Louvain-la-Neuve, Belgium
[d] Department of Materials Science. Polytechnic University of Madrid. E. T. S. de Ingenieros de Caminos. 28040 – Madrid, Spain

## Abstract

Precipitation during high temperature aging of Al−Cu alloys is analyzed by means of the integration of classical nucleation theory and phase-field simulations into a multiscale modelling approach based on well-established thermodynamics principles. In particular, thermal stability of $\theta''$, $\theta'$ and $\theta$ precipitates was assessed from first principles calculations of the Helmholtz free energy while homogeneous and heterogeneous nucleation of $\theta''$ and $\theta'$ was analysed using classical nucleation theory. Precipitate growth was finally computed by means of mesoscopic phase-field model. The model parameters that determine quantitatively the driving forces for each transformation were obtained by means of first principles calculations and computational thermodynamics. The predictions of the models were in good agreement with experimental results and provided a comprehensive understanding of the precipitation pathway in Al−Cu alloys. It is envisaged that the strategy presented in this investigation can be used in the future to design optimum microstructures based on the information of the different energy contributions obtained from first principles calculations.

---

[1] Corresponding Author.
Email address: javier.llorca@imdea.org (J. LLorca)

## 1. Introduction

Precipitation of intermetallic phases from supersaturated solid solutions by means of high temperature aging is the most common strategy to increase the strength of alloys [1−2]. Precipitation strengthening is particularly efficient in some metallic systems, such as Al- and Ni-based alloys, in which precipitation leads to the nucleation and growth of one or more metastable phases, which may co-exist with the thermodynamically stable forms of the phases. The overall strength of the alloy depends on the size, shape, spatial distribution and volume fraction of the different intermetallic phases [2−4]. The optimum heat treatments and alloy compositions have been obtained as a result of a detailed understanding of the precipitation processes and a lot of "trial and error" iterations.

A good example of this strategy can be found in the precipitation of Al−Cu alloys during high temperature aging. The experimental evidences was summarized in [5]. Al−Cu alloys containing up to ~5 wt. % of Cu aged above 180 ºC after the quenching from the single phase region of the phase diagram show three different types of precipitates, namely θ", θ' and θ. All of the them have the form of circular disks parallel to {100} planes of the fcc α-Al lattice. The unit cells of these precipitates and of the α–Al matrix are shown in Fig. 1. Among these three types of precipitates, both θ" and θ' are key strengthening phases. The θ" ($Al_3Cu$) precipitates have a face-centered tetragonal structure ($a_{\theta"}$ = 0.404 nm, $c_{\theta"}$ = 0.768 nm), which is obtained by replacing every fourth layer of Al atoms by a layer of Cu atoms on (001) planes while the same atomic structure is maintained (Fig. 1b). The orientation relationship between θ" and α–Al matrix are $(001)_{\theta"}//(001)_\alpha$ and $[100]_{\theta"}//[100]_\alpha$ and θ" precipitates have three orientation variants. The θ" precipitates are coherent disks with a $\{001\}_\alpha$ habit plane.

θ' ($Al_2Cu$) precipitates present a body-centered tetragonal structure ($a_{\theta'}$ = 0.404 nm, $c_{\theta'}$ = 0.580 nm). The $(001)_{\theta'}$ plane is similar to $(001)_\alpha$, except for the absence of an atom in the face center but the $(100)_{\theta'}$ and $(010)_{\theta'}$ planes are quite different from those of the α–Al lattice in terms of arrangement and distance between atoms (Fig. 1c). The orientation relationship between θ' precipitates and α–Al matrix is the same as that between θ" precipitates and α–Al matrix. θ' precipitates also have three orientation variants. The habit planes of θ' precipitates are the same as those of θ" and the broad faces of the plates are coherent with the Al matrix. However, the edges of the plates are semi-coherent. Finally, θ precipitates ($Al_2Cu$) are considered as the equilibrium phase and present a complex body-centered tetragonal structure ($a_\theta$ = 0.607 nm, $c_\theta$ = 0.487 nm) and a melting point of 590 ºC (Fig. 1d). This phase does not have lattice planes that match well those of the matrix and different orientation relationships and shapes has been observed [6−7]. Moreover, the matrix/precipitate interfaces are incoherent.

The kinetics of precipitate development has also been widely studied by means of transmission electron microscopy observations of aged samples [8−13] and, more recently, by means of in situ transmission electron microscopy [14]. It has been reported that the precipitation at high temperature begins with the homogeneous nucleation of the θ" phase. θ' precipitates tend to nucleate heterogeneously upon dislocations afterwards. As the aging time increases, the θ" precipitates grow and eventually disappear while θ' precipitates form. Finally, the θ phase may also nucleate on grain boundaries or at θ'/α–Al interfaces after still longer aging times.

This complex precipitation pathway can be rationalized by means of first-principles statistical mechanics approaches that link the electronic structure of the crystals with the thermodynamic and kinetic properties [15]. In particular, the formation enthalpy of the metastable phases can be obtained by means of first principles calculations, and the Helmholtz free energy can be determined as a function of temperature by including the vibrational entropic contribution. This information will be used to assess the thermodynamic stability of the different phases as a function of temperature. In addition, the critical size for the nucleation of each precipitate phase can be computed according to classical nucleation theory from the saddle point of the Gibbs free energy surface. To this end, the contributions from the chemical free energy, the elastic strain energy associated to the transformation strain and the interface energies were determined from computational thermodynamics and first principles calculations. The critical nucleus size determined from classical nucleation theory is then used as the starting point to simulate precipitate growth during aging by means of a novel approach based on the phase-field model, in which all the model parameters that control the free energy are also determined from computational thermodynamics and first principles calculations [16]. Overall, this investigation shows how classic nucleation theory and phase-field simulations can be integrated into a multiscale modelling approach (based on first principles calculations and computational thermodynamics data) to predict the size, shape and spatial distribution of the different metastable precipitates that develop during high temperature aging.

## 2. Thermal stability

First principles calculations are a reliable tool to analyse the thermal stability of intermetallic phases, θ", θ' and θ, that appear during high temperature aging of Al-Cu alloys [15, 17]. Stable phases are given by the minimum values of the Gibbs energy, $G$, which is expressed as:

$$G(p,T) = E + pV - TS = H(p) - TS(T), \quad (1)$$

where $E$ is the internal energy of the system, $p$ is the pressure, $T$ is the temperature, $S$ the entropy and $H$ the enthalpy. Under the assumption that the volume of the crystal does not change with temperature, this expression can be simplified to:

$$F(T) = E - TS(T) \quad (2)$$

where $F$ stands for the Helmholtz free energy (which will be referred from this point on as the free energy) and the comparison of the free energies of the different phases is sufficient to establish the phase stability. Under the assumption of constant volume, the influence of the thermal expansion in the free energy is neglected and, thus, the phonon frequencies are also independent from $T$. Thus, free energy of formation, $\Delta F$, of each phase with stoichiometry $Al_mCu_n$ can be obtained as:

$$\Delta F = \frac{F^{Al_mCu_n} - (mF^{Al} + nF^{Cu})}{m + n} \quad (3)$$

where $F^{Al_mCu_n}$ is the free energy of the phase and $F^{Al}$ and $F^{Cu}$ stand for the free energies of Al and Cu elements, respectively. The free energy $F^{Al_mCu_n}$ is given by:

$$F^{Al_mCu_n} = E^{Al_mCu_n} - TS^{Al_mCu_n}(T) \quad (4)$$

where $E^{Al_mCu_n}$ in the total energy of the phase at 0 K (referred from here on as the static energy) and $S^{Al_mCu_n}(T)$ is the entropic contribution that depends on the temperature $T$. Equivalent expressions are used for Al and Cu.

$E^{Al_mCu_n}$ at 0 K (as well as the energies of Al and Cu) can be determined by means of first principles calculations of the unit cell of each phase. Starting from the unit cell of each phase (Fig. 1), the structures were initially relaxed with regard to internal and external coordinates and the static ground state energy was obtained for all three compounds along with those of their constituent elements (Al and Cu) at 0 K from the relaxed structures. The computational details of the calculations are summarized in the Appendix 1.

The entropic contribution to the free energy in the case of strongly ordered intermetallic compounds mainly comes from the vibrational entropy, as the electronic and configurational entropies are considered to be relatively small [16]. The phonon contribution to the free energy can be accounted for by means of the quasi-harmonic approximation [18], where the thermal properties of solid materials are traced back to those of a system of non-interacting phonons whose frequencies are however allowed to depend on volume or on other thermodynamic constraints (more details are given in Appendix 1). The phonon entropic contribution can be expressed by [19]:

$$S(T) = k_B \int_0^\infty \frac{\frac{\hbar\omega}{k_B T}}{\exp\left(\frac{\hbar\omega}{k_B T}\right) - 1} g(\omega)d\omega - k_B \int_0^\infty g(\omega)\ln\left[1 - \exp\left(\frac{\hbar\omega}{k_B T}\right)\right]d\omega \qquad (5)$$

where $h$ is the reduced Planck's constant, $k_B$ the Boltzmann's constant, $\omega$ denotes the volume dependent phonon frequencies and g($\omega$) is the phonon density of states (DOS). The method of finite displacements (also known as the supercell method) [19] was used to determine the DOS of each phase and the details of the analyses can be found in Appendix 1.

The free energies of formation of θ'', θ' and θ phases are plotted as a function of temperature in Fig. 2. The enthalpies of formation at 0 K are similar to those reported in previous investigations [20−21] and that for θ compares well with the data in the literature to predict the experimental phase diagram of the Al-Cu system [22]. The effect of the temperature of the formation energy is very small for θ'', larger for θ and much larger for θ'. These two latter phases have the same stoichiometry (Al$_2$Cu) and their thermal stability as a function of temperature is given by the relative values of the formation free energy. The results in Fig. 2 indicate that θ' remains the stable phase up to 500 K, and transformation to θ is not expected even after very long aging times below this temperature. Although this is a generally unexpected result, it is also confirmed by the experimental work from Guinier *et al.* [23] who reported that θ' phase was the stable precipitate below 463 K. Moreover, recent differential scanning calorimetry analyses were also in agreement with these results [13]. Above that temperature, the vibrational contribution drives the free energy of θ below that of θ', and θ becomes the stable intermetallic phase at that stoichiometry. These results are in agreement with the work by Wolverton and Ozolins [20] that reported a transition temperature at ~423−473 K and Ravi and Ozolins [24] that reported that the θ' is the stable allotrope at 300 K.

The analysis of the stability of the θ'' phase with respect to θ'' and θ' has to be addressed using the convex hull diagram (Fig. 3) because of the different stoichiometry (Al$_3$Cu) of this compound. The data in Fig. 3 for 0 K and 528 K show that the entropic contribution fails to drive the free energy of formation of θ'' below the line connecting α-Al to θ' (low temperatures) and α-Al to θ (high temperatures) and, thus, θ'' is a metastable phase, in agreement with the accepted phase diagram [22].

## 3. Multiscale modelling of precipitate nucleation and growth

*3.1 Precipitate nucleation*

According to the classical nucleation theory, the nucleation energy barrier of a precipitate is given by the saddle point on the *ΔG* surface [25−26], which stands for the variation of the Gibbs free energy due to the formation of the precipitate. *ΔG* can be approximated by

$$\Delta G(A, V) = \Delta G_C V + G_S V + \Gamma \left[A, D\left(V^{\frac{2}{3}}\right)\right], \tag{6}$$

where *ΔG$_C$* is the chemical driving force for nucleation, $G_S$ stands for the elastic strain energy contribution, and *Γ* is the total interfacial energy. *ΔG* is a function of both the volume, *V*, and the aspect ratio, *A*, of the precipitate because both *ΔG$_S$* and *Γ* are functions of *A,* and *Γ* is also a function of the total interfacial area *D* which, in turn, depends on $V^{\frac{2}{3}}$.

The saddle point on *ΔG* surface that marks the nucleation energy barrier $\Delta G^*$ has to satisfy the conditions $\partial(\Delta G)/\partial V = 0$, $\partial(\Delta G)/\partial A = 0$, $\partial^2(\Delta G)/\partial A^2 > 0$ and $\partial^2(\Delta G)/\partial V^2 < 0$. To obtain an analytical form of each term in eq. (6), both θ'' and θ' precipitates are assumed to be oblate spheroids ($m_1 = m_2 > m_0$, where $m_1$, $m_2$ and $m_0$ are the length of the semi-axes of the spheroid) with an aspect ratio $A = m_1/m_0$, in agreement with recent in situ transmission electron microscopy observations [14].

*3.2 Precipitate growth*

Once a precipitate with a given size and aspect ratio has been nucleated, precipitate growth can be studied via the phase field method, in which the evolution of arbitrary morphologies can be simulated without explicitly tracking the interface [27−29]. In this method, the microstructure is described by the corresponding order parameters. For instance, a set of conserved and non-conserved order parameters, *x* and {$\eta_p$} are used to represent the spatial distribution of Cu concentration and the *p*th orientation variant of θ'' precipitate (*p* = ①, ② and ③) respectively. The corresponding phase field description of the order parameters for the θ' precipitate is given in [30].

The total free energy of the system, $G_{tot}$, is a function of these order parameters and can be expressed as:

$$G_{tot} = \int d\mathbf{r} \left[ G_C + \frac{\kappa_x}{2}(\nabla x)^2 + \frac{1}{2}\sum_{i=1}^{3}\sum_{j=1}^{3}\sum_{k=1}^{p} \beta_{ij}(p)\nabla_i \eta_k \nabla_j \eta_k \right] + G_S. \tag{7}$$

Thus, the energy contributions that control precipitate growth are the chemical free energy $G_C$, the elastic strain energy $G_S$ and the interfacial energy that is related to the gradient terms $\frac{\kappa_x}{2}(\nabla x)^2$ and $\frac{1}{2}\sum_{i=1}^{3}\sum_{j=1}^{3}\sum_{k=1}^{p}\beta_{ij}(p)\nabla_i\eta_k\nabla_j\eta_k$, where $\kappa_x$ and $\eta_p$ are gradient coefficients of *x* and {$\eta_p$}, respectively. These gradient terms represent the energies associated with the variation of *x* and {$\eta_p$} at the α–Al/θ'' or α–Al/θ' interfaces. Both the interfacial energy and the interfacial width are related to the gradient coefficients. If the system contains only one order parameter, the dependence of the interfacial energy and of the interface thickness with the order parameters can be expressed analytically [31]. If the system is described by several order parameters (as in this case), the interfacial energy and interfacial thickness have to be determined numerically by solving the Cahn-Hilliard and the Allen-Cahn equations. It is generally accepted that the interfacial energy and the width of the profile of the order parameters increase with the gradient energy coefficients. If the set of phase-field variables contains composition fields, only the gradient energy coefficients have to be adjusted to reproduce the interfacial energy and the width [31–32].

Within the phase-field framework, precipitate growth is simulated through the evolution the order parameters as a function of time $t$, according to the Cahn-Hilliard and Allen-Cahn equations [33–34]:

$$\frac{1}{V_m^2}\frac{\partial x}{\partial t} = \nabla \cdot \left[M \nabla \left(\frac{\delta G_{tot}}{\delta x}\right)\right] \tag{8}$$

$$\frac{\partial \eta_p}{\partial t} = -L \frac{\delta G_{tot}}{\delta \eta_p}, \tag{9}$$

where $M$ and $L$ are the chemical and interface mobility coefficients and $V_m$ is the molar volume. They were chosen as $V_m^2 M = 1$ and $L = 5$, following previous simulations of the growth of θ' [30]. The relationship between the dimensionless parameters in the phase field model and the dimensional physical parameters are shown in Appendices 3 and 4 for θ", while those for θ' can be found in [30].

In the following subsections, the expressions to determine the contributions coming from the chemical free energy, the interface energy and the elastic strain energy used in the nucleation and phase field models are detailed. In addition, the elastic interaction energy term that arises from the interaction of the strain field around the precipitate with other strain fields (induced by dislocations, other precipitates, and external stresses) is also presented.

*3.3 Chemical free energy*

The chemical Gibbs free energy of the α–Al matrix, $f_\alpha$, is a function of the Cu concentration $x$ for a given temperature and can be obtained from the CALPHAD database [35]. From this information, the chemical driving force for the nucleation of a precipitate, $\Delta G_C$, can be expressed as:

$$\Delta G_C = -[f'_\alpha(x_0) - f'_\alpha(x_{eq})]x_p - f_\alpha(x_0) + f_\alpha(x_{eq}) + f'_\alpha(x_0)x_0 - f'_\alpha(x_{eq})x_{eq} \tag{10}$$

where $f'_\alpha$ stands for the derivative of chemical Gibbs free energy of the α–Al matrix and its derivative and $x_0$ and $x_{eq}$ are the initial and equilibrium Cu concentration in α–Al matrix at the aging temperature, respectively. $x_p$ is the Cu concentration in the precipitate. he detailed derivation of eq. (10) can be found in Appendix 2. Please note that $\Delta G_C$ in eq. (10) is expressed in J/mol, and has to be transformed to J/m³ when substituted into eq. (6). Assuming an initial Cu content in the α–Al matrix of 4 wt. % ($x_0 = 0.0174$) and an aging temperature of $T = 180$ °C (453 K) [22], $x_{eq}$ is ≈ 0.001 according to the available Al-Cu phase diagram. Then, the chemical driving forces for the nucleation of the θ" and θ' precipitates, $\Delta G_C^{\theta''}$ and $\Delta G_C^{\theta'}$, can be estimated, through eq. (10), as $\Delta G_C^{\theta''} = -1.00 \times 10^8$ J/m³ and $\Delta G_C^{\theta'} = -1.28 \times 10^8$ J/m³.

The chemical free energy contribution to the θ" precipitate growth, $G_C$, is expressed as a function of order parameters and of the chemical Gibbs free energies of α–Al and θ" phases, i.e., $f_\alpha$ and $f_{\theta''}$:

$$G_C(x, \{\eta_v\}) = \left[ f_\alpha \left( 1 - \sum_{i=1}^{p} H(\eta_i) \right) + f_{\theta''} \sum_{i=1}^{p} H(\eta_i) \right] + B \sum_{i=1}^{p} \sum_{j \neq i}^{p} [\eta_i^2(\mathbf{r}) \eta_j^2(\mathbf{r})]. \quad (11)$$

Since the chemical Gibbs free energy of the metastable phase θ" is not available, $f_{\theta''}$ is approximated by a 4$^{th}$ order parabolic function (see Appendix 3 for details). $H(\eta_i)$ is an interpolation function, which is used to connect $f_\alpha$ and $f_{\theta''}$, and which has the form $H(\eta_i) = \eta_i^3(10 - 15\eta_i + 6\eta_i^2)$. The last term in eq. (11) is added in order to avoid the coalescence of different θ" variants and $B$ was set to 10 according to the domain wall energy between different θ" variants [29]. Equivalent expressions are used for the θ' precipitate and they can be found in [30].

*3.4 Interfacial energy*

Both θ' and θ" precipitates have three orientation variants due to the cubic symmetry of the fcc α–Al lattice. Nuclei of θ" and θ' belonging to the variant ① were chosen for the analysis in the followings. The orientation relationships between this variant of θ' and θ" precipitates and the α–Al are (001)$_{prep}$//(001)$_\alpha$ and [100]$_{prep}$//[100]$_\alpha$ [5, 11−13, 28], leading to two major different types of possible interfaces for each precipitate. In the case of θ', the (001)$_{θ'}$/(001)$_\alpha$ interface is coherent while both (100)$_{θ'}$/(100)$_\alpha$ and (010)$_{θ'}$/(010)$_\alpha$ interfaces are semi-coherent [5, 36]. Two different types of interfaces were also found in the θ" precipitate, although both are coherent with the matrix (Fig. 4). The atomic structures of α–Al and θ" along the (001)$_{θ''}$/(001)$_\alpha$ interface (type $I_1$ interface, Fig. 4a) were more similar than along the (100)$_{θ''}$/(100)$_\alpha$ and (010)$_{θ''}$/(010)$_\alpha$ interfaces (type $I_2$ interface, Fig. 4b). The interfacial energies of the four types of interfaces between θ' and θ" precipitates and the α−Al matrix were determined by means of first principles calculations using the methodology presented in [13] and the corresponding values are summarized in Table 1. They show that the energy of the (001)$_{θ''}$/(001)$_\alpha$ interface, i.e., $\gamma^{\theta''}_{(001)}$, is lower than that of $\gamma^{\theta''}_{(100)}$ and $\gamma^{\theta''}_{(010)}$ for each precipitate. Thus, according to the Wulff plot, the minor axes of the θ" spheroid was parallel to [001]$_{θ''}$ because the precipitate/matrix interfacial energy in the plane perpendicular to this direction, $\gamma^{\theta''}_{(001)}$, is minimum. The two major axes were parallel to [100]$_{θ''}$ and [010]$_{θ''}$. Similar orientations were obtained for the variant ① of a θ' nucleus following the same reasoning. For these orientations, the total interfacial energy of an oblate nucleus is given by

$$\Gamma = \int_0^{\frac{\pi}{2}} (4\pi)^{1/3} \left( \frac{3V}{A^2} \right)^{2/3} \cos t \sqrt{A^2 \left[ \gamma^{prep}_{(001)} \right]^2 \sin^2 t + \left[ \gamma^{prep}_{(010)} \right]^2 \cos^2 t} \, dt \quad (12)$$

where *t* is an integration parameter.

In phase field method, the interfacial energy is described as the excess energy due to inhomogeneities in the system [31]. The interfacial energy consists of two contributions: gradient in the field variable and bulk free energy from material in the interface. The interfacial energy is related to the gradient terms in eq. (7) and the anisotropic interfacial energy is incorporated in the third rank tensor $\beta_{ij}(p)$. For variant ① of θ", $\beta_{ij}(p)$ is given by:

$$\beta_{ij}(①) = \begin{pmatrix} 12 & 0 & 0 \\ 0 & 12 & 0 \\ 0 & 0 & 0.15 \end{pmatrix} \quad (14)$$

and the values in eq. (14) were chosen together with the gradient coefficient of the concentration field, $\kappa_x = 0.6$, to ensure that the interfacial energies of different types of interfaces are consistent with first principles calculation results and to avoid artificial fraction.

*3.5 Elastic strain energy*

In both nucleation and growth stages, the elastic strain energy was calculated using Eshelby's model [37−39]. The precipitate nucleus is assumed to have the shape of an oblate spheroid and the corresponding elastic strain energy is given by [39−40]:

$$G_S = -\frac{1}{2} C_{ijkn}\left(\varepsilon_{kn} - \varepsilon_{kn}^p\right)\varepsilon_{ij}^p, \quad (15)$$

where $C_{ijkn}$ is the elastic stiffness tensor. It is assumed, for simplicity, that the elastic constants of the Al matrix and the precipitate phases are the same ($C_{11}$ = 110.4 GPa, $C_{12}$ = 60.0 GPa and $C_{44}$ = 31.6 GPa [24]). $\varepsilon_{kn}$ is the constrained strain, which is related to the stress-free transformation strain (SFTS) of the nucleus, $\varepsilon_{ij}^p$, by

$$\varepsilon_{ij} = W_{ijkl} C_{klmn} \varepsilon_{mn}^p, \quad (16)$$

where $W$ is the Eshelby's tensor. The directions $[100]_\alpha$, $[010]_\alpha$ and $[001]_\alpha$ are chosen parallel to the $x$, $y$ and $z$ axes, respectively, of the reference Cartesian coordinates, with the minor axis of the oblate spheroidal nucleus parallel to the $z$ axis and the two major axes located in the $xy$ plane. $W$ is expressed as [40]:

$$W_{ijkl} = \frac{1}{8\pi} \int_0^{2\pi} d\rho \int_0^\pi A \sin\varphi \, d\varphi \frac{z_i z_l N_{jk}^{-1} + z_j z_l N_{ik}^{-1}}{(\sin^2\varphi + A^2 \cos^2\varphi)^{3/2}}, \quad (17)$$

$$N_{ik} = C_{ijkl} z_j z_l, \quad (18)$$

$$\mathbf{z} = [\sin\varphi \cos\rho, \sin\varphi \sin\rho, \cos\varphi]. \quad (19)$$

In eqs. (16) – (18), $\rho$ is an integration parameter, $\varphi$ the angle between the vector $z$ and the minor axis of the oblate spheroid nucleus, and $C_{ijkl}$ is referred to a Cartesian coordinate basis whose axes are parallel to the oblate spheroid. The SFTS, $\varepsilon_{ij}^p$, is obtained from the transformation matrix $T_{prep}$ as:

$$\varepsilon_{ij}^p = \frac{T_{prep}^T T_{prep} - I}{2}, \quad (20)$$

where $T^T$ is the transpose matrix of $T$ and $I$ is the identity matrix. The forms of $T$ for the θ" and θ' precipitates belonging to the variant ① are given by [28, 30, 41]:

$$T_{①}^{\theta''} = \begin{pmatrix} \dfrac{a_{\theta''}}{a_\alpha} & 0 & 0 \\ 0 & \dfrac{a_{\theta''}}{a_\alpha} & 0 \\ 0 & 0 & \dfrac{c_{\theta''}}{2a_\alpha} \end{pmatrix}, \quad T_{①}^{\theta'} = \begin{pmatrix} \dfrac{a_{\theta'}}{a_\alpha} & 0 & 0 \\ 0 & \dfrac{a_{\theta'}}{a_\alpha} & -\dfrac{1}{3} \\ 0 & 0 & \dfrac{2c_{\theta'}}{3a_\alpha} \end{pmatrix}, \quad (21)$$

where $a_{\theta''}$, $c_{\theta''}$; $a_{\theta'}$, $c_{\theta'}$; and $a_\alpha$ are lattice parameters of θ", θ' and α–Al matrix. The transformation matrices of the other variants of θ" and θ' can be obtained via symmetry operations.

For the simulation of the precipitate growth, the elastic energy contribution has to take into account the diffuse nature of the interface in the phase-field model and that the precipitate shape may differ from an ellipsoid. Thus, $G_S$ in eq. (7) is given in a more general form as a function of order parameter based on Khachatruyan and Shatalov's microelasticity theory [38]:

$$G_S = \frac{1}{2} \sum_{p=1} \sum_{q=1} \int B_{pq}\left(\frac{\mathbf{g}}{|\mathbf{g}|}\right) \{\eta_p(\mathbf{r})\}_\mathbf{g} \{\eta_q(\mathbf{r})\}_\mathbf{g}^* \frac{d^3\mathbf{g}}{(2\pi)^3}, \quad (22)$$

where $B_{pq}\left(\dfrac{\mathbf{g}}{|\mathbf{g}|}\right)$ is given by

$$B_{pq}\left(\frac{\mathbf{g}}{|\mathbf{g}|}\right) = \begin{cases} 0 & \mathbf{g} = 0 \\ C_{ijkl}\varepsilon_{ij}^p(p)\varepsilon_{kl}^p(q) - n_i\sigma_{ij}(p)\Omega_{jk}\sigma_{kl}(q)n_l & \mathbf{g} \neq 0 \end{cases}. \quad (23)$$

The integral in eq. (22) is taken in the reciprocal space and $\mathbf{g}$ is a vector in the reciprocal space. Note that $\mathbf{g} = 0$ is excluded from the integration, which defines the principal value. $\{\eta_p(\mathbf{r})\}_\mathbf{g}$ is the Fourier transform of η(**r**) and the superscript * indicates the complex conjugate. $\mathbf{n} = \mathbf{g}/|\mathbf{g}|$ is a unit vector in reciprocal space, $\Omega_{jk}^{-1} = C_{ijkl}n_k n_l$ and $\sigma_{ij}(p) = C_{ijkl}\varepsilon_{kl}^p(p)$.

*3.6 Elastic interaction energy*

The strain field induced by the transformation of the matrix to the precipitate may interact with other pre-existing stress fields in the matrix, $\sigma_{ij}^0(\mathbf{r})$, due to the presence of dislocations or of other precipitates. The influence of the pre-existing stress field on the nucleation and growth of the precipitate can be assessed by including an extra elastic interaction energy term, $G_I$, in eqs. (6) and (7). This contribution can be expressed as [38−40, 42−44]

$$G_I = -\sigma_{ij}^0(\mathbf{r}) \sum_{p=1}^n \varepsilon_{ij}(p) V = -\sigma_{ij}^0(\mathbf{r}) \int \sum_{p=1}^n \varepsilon_{ij}(p)\eta_p(\mathbf{r}) \, d\mathbf{r}. \quad (24)$$

The contribution of the pre-existing stress field, $\sigma_{ij}^0(\mathbf{r})$, to the formation of the *p*th variant of the precipitates can be evaluated as the variational derivative of $G_I$ to η,

$$g_I = \frac{\delta G_I}{\delta \eta_p} = -\sigma_{ij}^0(\mathbf{r})\varepsilon_{ij}(p). \quad (25)$$

When different variants of the to-be-nucleated precipitates form under the influence of the pre-existing stress fields, $g_I$ depends on the SFTS of each variant. The variant with the lowest (more negative) $g_I$ will stand for the most energetically favoured variant to be nucleated.

## 4. Precipitate nucleation

*4.1 Homogeneous nucleation*

Homogeneous nucleation of the θ" and θ' precipitates was analysed from the variation of the Gibbs free energy according to eq. (6). All contributions coming from the chemical free energy, the elastic strain and the interfacial energy increase with the volume of the precipitate but the two latter also depend on the aspect ratio of the nucleus according to eqs. (12) and (16). This dependence is more complex and can be analysed independently assuming that the volume of both θ" and θ' nuclei is set to 1, i.e., the unit volume. In this case, the calculated interfacial energies of θ" and θ' nuclei are shown in Figs. 5a and 5b, respectively, as a function of the aspect ratio $A$. According to Fig. 5a, the interfacial energy between the θ" precipitate and the α-Al matrix, $\Gamma_{\theta"}$, decreases with the aspect ratio until $A = 3.70$ and then increases with $A$. This result is consistent with the Wulff plot because the ratio between $\gamma_{(010)}^{\theta"}$ (~23.3 mJ/m$^2$) and $\gamma_{(001)}^{\theta"}$ (~6.3 mJ/m$^2$) is also ~3.69. Similarly, the minimum interfacial energy between the θ' precipitate and the α-Al matrix, $\Gamma_{\theta'}$, is attained for $A = 3.2:1$ (Fig. 5b), which is equal to the ratio between $\gamma_{(010)}^{\theta'}$ (487 mJ/m$^2$) and $\gamma_{(001)}^{\theta'}$ (152 mJ/m$^2$). It should be noted that the minimum value of $\Gamma_{\theta"}$ (~0.073 mJ/m$^2$) was smaller than that of $\Gamma_{\theta'}$ (~1.598 mJ/m$^2$).

Similarly, the elastic strain energies of the θ" and θ' nuclei are plotted in Figs. 5c and 5d, respectively, as a function of aspect ratio $A$. The elastic strain energy decreased as the aspect ratio of the nuclei increased, showing that a nucleus is favoured to maximise/minimise its length along the direction that has the minimum/maximum transformation strain from the viewpoint of the elastic strain energy. In the case of θ", the transformation strain of variant ① perpendicular to the $(001)_\alpha$ plane is maximum (−5.4%), and those in the $(001)_\alpha$ plane along the $[100]_\alpha$ and $[010]_\alpha$ directions are minima (−1.7%). In the case of θ', the $(001)_\alpha$ is an invariant plane of variant ① during the α–Al → θ' phase transformation [29]. Thus, the elastic strain energy associated with the SFTS favours the development of an oblate ellipsoids with minor axes along $[001]_\alpha$ and the two major axes in the plane $(001)_\alpha$ for both θ" (variant ①) and θ' (variant ①) nuclei.

The variation in the Gibbs free energy due to the nucleation of the θ" precipitate, $\Delta G$, is plotted as a function of $V$ and $A$ in Fig. 6a and the nucleation energy barrier, $\Delta G^*$, is given by the saddle point on the $\Delta G$ surface. The 2D projection of the $\Delta G$ surface is shown in Fig. 6b. The black line and arrow Fig. 6b reveal the minimum energy path for the nucleation [45], which is given by the locus on the $\Delta G$ surface that satisfies $\partial(\Delta G)/\partial A = 0$ as the precipitate volume increases. The energy barrier for homogeneous nucleation of θ" is $\Delta G^*_{\theta"} = 2.48 \times 10^{-20}$ J, while the corresponding critical volume and aspect ratio are $V^*_{\theta"} = 0.95$ nm$^3$ and $A^*_{\theta"} = 6.7$, respectively (Table 2). The critical aspect ratio is higher than that predicted by the Wulff plot because of the contribution of the elastic strain energy. In fact, if this term is not taken into account in eq. (6), the critical aspect ratio of the θ" precipitate nucleus was equal to 3.7, the ratio between the interface energies $\gamma_{(010)}^{\theta"}$ and $\gamma_{(001)}^{\theta"}$.

The nucleation energy barrier of θ' was studied following the same procedure and the corresponding $\Delta G$ surface is plotted in Fig. 7a while the 2D projection of the $\Delta G$ surface is shown in Fig. 7b. The nucleation energy barrier for θ' is $\Delta G^*_{\theta'} = 8.30 \times 10^{-16}$ J, four orders of magnitude higher than that of θ". In addition, the volume of the critical nucleus of θ', $V^*_{\theta'} = 59365$ nm$^3$, is two orders of magnitude larger than that of θ" (Table 2). Therefore, the homogeneous nucleation of θ' is much more difficult than that of θ", in agreement with the experimental observations. It is worth noting that the critical aspect ratio $A^*_{\theta'} = 36$ is much higher than the ratio between the energies of the semi-coherent ($\gamma^{\theta'}_{(010)}$) and coherent ($\gamma^{\theta'}_{(001)}$) interfaces (487/152 = 3.2) (Table 1). This result indicates that the contribution of the elastic strain energy plays a dominant role (in comparison with the interface energies) in the nucleation of θ'.

*4.2 Heterogeneous nucleation*

*4.2.1 Heterogeneous nucleation of θ"*

Previous experimental results show that θ" precipitates are randomly distributed in α−Al matrix [4, 12]. These θ" may form on pre-existing Guinier-Preston zones and this phenomenon has been previously studied using first principles calculations [46]. Nevertheless, dislocations may also act as heterogeneous nucleation sites. In this section, the effect of a pre-existing dislocation loop in a $\{111\}_\alpha$ plane with a Burgers vector $a_\alpha \langle \bar{1}10 \rangle / 2$ (Fig. 8a) on the nucleation of θ" precipitates was investigated from the magnitude of the $g_I$ in eq. (24) due to the interaction of the stress field of the dislocation loop and the SFTS of the to-be-nucleated θ" variant.

As described in section 3.4, nucleation is most favoured at the positions where $g_I$ is minimum (most negative) and the extra driving force for the formation of the first nucleus of the precipitate on the dislocation can be simplified to $G_I = g_I^m V$, where $g_I^m = \min(g_I)$. Fig. 8b shows the distribution of the interaction energy in a (111) plane that is 0.25 nm (around one Burgers vector) below the dislocation loop. The interaction energy in this plane is negative on the left-hand side of the dislocation loop, and the minimum value $e_{int}^m$ is marked by Q in Fig. 8b. Thus, θ" precipitates are most favoured to form at Q, which stands below the edge-like region of the dislocation loop. Similar conditions were found above the dislocation loop and they are not discussed further. Moreover, the positions corresponding to $g_I^m$ for other precipitate variants are located at the same position and are not shown for the sake of brevity. The min($g_I$) values are negative for all three variants of θ", i.e., the stress field of the dislocation loop facilitates the nucleation of precipitates. When θ" forms on the dislocation loop, min($g_I$) values of variants ② and ③ are approximately −0.105 and −0.095 × 10$^7$ J/m$^3$, respectively, which are slightly larger than that of variant ① (around −0.089 × 10$^7$ J/m$^3$).

The $\Delta G$ surface for heterogenous nucleation of a θ" precipitate (variant ①) under the influence of the pre-existing dislocation loop shown in Fig. 8a is plotted in Fig. 9a. The comparison of the $\Delta G$ surfaces corresponding to the homogeneous (Fig. 6a) and heterogeneous (Fig. 9a) nucleation of θ" shows that the presence of the dislocation reduces the nucleation energy barrier (given by the saddle point P in both figures) from ~2.48 × 10$^{-20}$ J to ~1.75 × 10$^{-20}$ J, and the critical volume from ~0.95 nm$^3$ to ~0.56 nm$^3$ (Table 2). Thus, the energy barrier and the critical size for the heterogeneous nucleation of θ" precipitates on dislocations are lower than those for homogeneous nucleation. However, it should be noted that the lattice parameters and the volume of the precipitates vary in discrete steps during nucleation and growth. This discreteness was not included in the calculation of the $\Delta G$ surfaces and this difference may influence the accuracy of the prediction of $\Delta G^*$ in the case of θ". For example, the critical thickness of homogeneously and heterogeneously nucleated θ" nuclei are 0.34 nm and 0.31 nm, respectively. Both values are smaller than the lattice parameter of θ" along $c$ direction (0.77 nm), which is the minimum thickness of θ". Thus, the calculated critical thickness of the nucleus cannot be reached in the case of θ" precipitates, and this limitation of the model indicates that the influence of the dislocation loop on the nucleation of θ" may be weaker in the actual material than in the model because the presence of the dislocation cannot reduce further the critical thickness necessary for the homogeneous nucleation of the precipitate.

*4.2.2 Heterogeneous nucleation of θ'*

Some previous experimental evidences and calculation results indicate that θ' precipitates may nucleate on dislocations and/or θ" precipitates [5, 13, 46]. Both scenarios are analysed below.

*4.2.2.1 Heterogeneous nucleation of θ' on a dislocation loop*
The $ΔG$ surface of a θ' nucleus (variant ①) under the influence of the stress field of the same dislocation loop is calculated and the result is plotted in Fig. 9b. Compared with the value of $ΔG^*$ shown in Fig. 9b and that in Fig. 7a, it is found that the presence of the dislocation loop reduces the nucleation energy barrier by 88% (from ~$2.1 \times 10^{-15}$ J, Fig. 7a, to $2.5 \times 10^{-16}$ J, Fig. 9b) and the critical volume by one order of magnitude (from ~$5.93 \times 10^4$ nm$^3$, Fig. 7a, to ~$4.31 \times 10^3$ nm$^3$, Fig. 9b) (Table 2).

The $g_I^m$ values for interaction energy between a θ' precipitate and an edge dislocation, a screw dislocation and a dislocation loop [28] are compared in Fig. 10 with those corresponding to θ" precipitates. $g_I^m$ for θ" precipitates are far less negative than those for θ' when the precipitates form on any type of dislocation (either edge, screw or a loop). In addition, the nucleation energy barrier of a θ' precipitate is reduced in more than order of magnitude in the presence of a dislocation loop (Fig. 9b). These results indicate that compared with θ", θ' is more favoured to nucleate and grow on pre-existing dislocations in terms of elastic strain energy. This is proved by previous simulation results [30, 47] and experimental observations [5, 13, 48], which show that θ' is favoured to nucleate on and grow along the pre-existing dislocation lines, and the shape of the heterogeneously formed precipitates is no longer circular.

*4.2.2.2 Heterogeneous nucleation of θ' on a θ" precipitate*
Pre-existing θ" precipitates may also act as heterogeneous nucleation sites of θ'. To study whether nucleation of θ' is favoured on pre-existing θ", the interaction energy density, $g_I$, fields between the stress field of a pre-existing θ" and the SFTS of 12 deformation variants of θ' are calculated and the results are shown in Fig. 11. In each plot, a pre-existing θ" precipitate (variant ①) with a ~20 nm diameter and a ~ 2.5 nm in thickness is shown at the centre. θ" precipitates of these dimensions are commonly observed in experiments [5, 13]. The blue regions around the θ" precipitates represent the zones where $g_I < −3$ MJ/m$^3$, which are the possible heterogeneous nucleation sites of θ'. It is worth noting that the habit planes of the variants ①－④ of the to-be-nucleated θ' are parallel to that of the pre-existing θ" precipitate while the habit planes of other variants are perpendicular to that of the θ" precipitate. The blue regions do not appear in Figs. 11f, h, j, l, indicating that the formation of variants ⑥, ⑧, ⑩, ⑫ of θ' is not favoured by the presence of the θ" precipitate. All other variants are favoured to nucleate on the rim of the pre-existing θ" precipitate but the heterogeneous nucleation sites depend on the particular variant considered. Both variants ① and ⑦ (Figs. 11a and g) favour the nucleation on the upper left and lower right sides of the θ" precipitate, while variants ③ and ⑤ (Figs. 11c and e) prefer to form on the lower left and upper right sides. Similarly, variants ②, ⑪ (Figs. 11b and k) and ④, ⑨ (Figs. 11d and i) are favoured to nucleate on the opposite sites of the pre-existing θ" precipitates as well.

The minimum interaction energies, $g_I^m$, between the stress-field of the θ" precipitate and the SFTS of the to-be-nucleated θ' variants are shown in Fig. 12. $g_I^m$ values corresponding to variants ⑥, ⑧, ⑩, ⑫ of θ' are much less negative than those of other variants, which is consistent with the previous plot. The values of $g_I^m$ corresponding to the other 8 variants of θ' are nearly the same and around −4 MJ/m$^3$. By incorporating the $g_I^m$ contribution to eq. (6), the $\Delta G$ surface of a θ' (variant ①) on a pre-existing θ" precipitate can be calculated and is shown in Fig. 13. The values of $\Delta G_{\theta'}^*$ and $V_{\theta'}^*$ are ~7.0 × 10$^{-16}$ J and ~1.41 × 10$^4$ nm$^3$, respectively (Table 2). They are much lower than those calculated for homogeneous nucleation of θ' precipitates (Fig. 7a) but still higher than those for heterogeneous nucleation of θ' precipitates on dislocations (Fig. 9b and [30]). This is reasonable because in case of the nucleation of θ' on pre-existing θ" precipitates, $g_I^m$ is nearly half of the $g_I^m$ value when θ' forms on pre-existing dislocations.

Besides heterogeneous nucleation on dislocations and θ" precipitates, θ' may also directly transform from θ" [14]. In principle, this *in situ* transformation can be analysed using the same methodology proposed above to determine the nucleation driving force. Nevertheless, this task will require to know the details of the transformation to determine the SFTS as well as the interfacial energies and this information is not available.

Finally, the limitations associated with the classical nucleation theory to simulate precipitate nucleation should be indicated. The first one comes from the continuum nature of the approach, in which atomic fluctuations are not considered. In comparison, first principles calculations can be used to study nucleation of very small precipitates (for instance, Guinier Preston zones or very small θ" precipitates) taking into account atomic fluctuation. Nevertheless, this strategy cannot be extended to analyse nucleation when the critical nucleus volume is very large (> 500 nm$^3$) because of the limitations imposed by the atomistic calculations. In this regime, the statistical effects due to atomic fluctuations are more limited and classical nucleation theory can provide accurate predictions.

Another limitation from classical nucleation theory comes from the fact that the transition states, that dictate the minimum energy path for the transformation, have to be provided by means of atomistic simulations or experimental observations. In this respect, application of classical nucleation theory to simulate the direct transformation θ" into θ' is hindered by the lack of detailed information about the transition states during the transformation.

In summary, the nucleation analysis based on first principles calculations explains the precipitation sequence observed in Al-Cu alloys aged at high temperature (around 180 ºC). Precipitation begins with the homogeneous nucleation of θ" precipitates. This is followed by heterogeneous nucleation of θ' precipitates firstly on pre-existing dislocation and later on θ" precipitates. Further ageing will lead to the progressive transformation of θ' into θ" precipitates, because the latter are metastable, and to the homogeneous nucleation of θ' precipitates.

## 5. Precipitate growth

The growth of precipitates was simulated using a mesoscale phase field model [29−30]. Simulations were carried in a cubic box of $256l_0 \times 256l_0 \times 256l_0$ for an Al−Cu alloy aged at 180 °C. According to the interfacial energies in Section 2, the grid spacing of the phase-field simulation was $l_0 = 0.5$ nm, and thus, the cell dimensions were $128 \times 128 \times 128$ nm$^3$. An initial precipitate (either θ' or θ") was located at the center of the cubic box. The initial size and shape of the precipitate was given by critical values of the nucleation volume ($V^*$) and aspect ratio ($A^*$) in Table 2. However, in some cases, the critical thickness of the nucleus, such as homogeneous nucleation of θ" precipitates, was smaller than the grid size of the cell. In this case, the minimum energy path for nucleation in Fig. 6b was followed until the thickness of the precipitate in the path matches the grid size. This precipitate is then introduced into the cubic box to begin the phase-field simulations. To determine the precipitate size, it is assumed that a grid point belongs to the $p$th variant of the precipitate if $\eta_p > 0.5$. Otherwise it is assumed to belong to the matrix.

The evolution of the shape of a θ" precipitate (orientation variant ①) is shown in Figs. 14a to 14c as a function of the simulation time t*. The initial θ" precipitate (Fig. 14a) was obtained by the classical nucleation theory and had the shape of an oblate ellipsoid with $(001)_\alpha$ habit plane and an aspect ratio of ~9. The precipitate kept the same habit plane during growth but the aspect ratio increased up to ~13 when the thickness of the precipitate was ~6 nm (Fig. 14c). It should be noted that both the interfacial and the elastic strain energies promoted the growth of the orientation variant ① of θ" precipitate in the $(001)_\alpha$ plane because the $(001)_{\theta"}/(001)_\alpha$ interface has the lowest interfacial energy and the large transformation strain perpendicular to the $(001)_{\theta"}$ plane hinders the growth of θ" precipitate in this direction.

The progressive increase in the aspect ratio of the precipitate was due to the interplay between the anisotropic interface energy and the elastic strain energy. If the phase-field simulations only included the contribution of the chemical free energy and of the anisotropic interfacial energy, the aspect ratio of the precipitate remained constant and equal to 3.70, which is equal to ratio in the interfacial energy between $(100)_{\theta"}/(100)_\alpha$ and $(001)_{\theta"}/(001)_\alpha$ interface and in agreement with Wulff plot. The elastic strain contribution tends to increase the aspect ratio and is proportional to the volume of the precipitate while the interface energy contribution increases with the precipitate area. As a result, the aspect ratio of the θ" precipitate increased with size.

The evolution of the shape of the θ' precipitate (variant ① in [30]) is shown in Figs. 14d to 14f. The grid size was changed to 0.8 nm in these simulations to represent better the evolution of the shape of the θ'. The initial precipitate size and shape were given by the classical nucleation theory (Table 2, homogeneous nucleation of θ', aspect ratio ~36) and were much larger than those of the θ" precipitate. The θ' precipitate grew along the same $(001)_\alpha$ habit plane and the aspect ratio sis not change significantly during growth (Fig. 14f), leading to a disk-shaped precipitate. It should be noted, however, that the main contribution to the high aspect ratio of the θ' precipitate came from the large shear deformation (1/3) associated with the SFTS, while the anisotropy in the interfacial energies (Table 1) was minor in this case.

The spatial distribution of θ" precipitates during homogeneous nucleation was simulated within the framework of phase-field approach using the explicit nucleation method [49] in which the nucleation rate of θ" was controlled by the nucleation energy barrier, $\Delta G$ calculated in Section 3. The simulation results are depicted in Fig. 15a, which shows the random distribution of 11 θ" precipitates. The precipitate colour indicates the variant and the habit planes of variants ①, ② and ③ are $(001)_\alpha$,

(010)$_\alpha$ and (100)$_\alpha$, respectively, which are coloured in red, blue and green. These three variants can be observed in Figs. 15b, c and d in detail. Notice that the broad faces of the precipitates are nearly equiaxed. The diameter and thickness of these precipitates were in range of 10 to 30 nm and 1.5 to 4.0 nm, respectively. These results are consistent with the previous analyses based on the interplay between anisotropic interfacial and elastic strain energies, and previous experimental observations [5, 13].

The equilibrium shape of the θ" and θ' precipitates (as given by the ratio between the average diameter and the thickness) was computed for precipitates of different size by changing the initial composition of the supersaturated solid solution. Only one precipitate was contained in the simulation cell in this case for convenience. The average precipitate diameter and thickness are plotted in Fig. 16 together with experimental data for θ" precipitates [13]. The experimental results were obtained in an Al−1.74 at. % Cu alloy aged at 180 °C for 18, 30 and 120 hours. The precipitate diameter and thickness were measured by transmission electron microscopy in thin foils extracted from the aged samples using the methodology presented Nie and Muddle [10]. The dimensions of 50 to 100 precipitates were measured in the samples aged for different times to get the average values reported in Fig. 16. For the sake of completion, the simulation results using the same approach and the experimental data for θ' precipitates in obtained using the same experimental methodology are included in Fig. 16 as well. The agreement between the numerical predictions and the experimental observation for the diameter and thickness of the θ" precipitates is good for precipitates whose diameter is in the range 10 to 50 nm. The error bars in the values of the precipitate thickness obtained from the phase-field model reflect the diffuse nature of the interface in the phase-field model and the grid size.

It should be emphasized the multiscale nature of the approach: all the parameters in the phase-field model were obtained from first-principles calculations and CALPHAD databases and there are not empirical fitting parameters. In addition, it should also be noticed that the thickness of the smallest θ" precipitates in the simulations (~1.25 nm) is close to two times the lattice parameter along the *c* axis (0.77 nm), pushing the continuum phase-field model to the limit.

## 6. Conclusions

The main phenomena that determine the precipitation pathway (thermal stability, nucleation and growth of precipitates) during high temperature aging of an Al-Cu alloys have been studied by means of the integration of classic nucleation theory and phase-field simulations into a multiscale modelling approach. The analysis is based on well-established thermodynamics principles and the parameters that determine the driving forces for each type of precipitate and process were obtained by means of first principles calculations and computational thermodynamics, leading to predictions that are independent of adjustable coefficients.

The stability of θ", θ' and θ phases was assessed from the free energy of formation, which was determined as a function of temperature by means of first principles calculations, including the contribution of vibrational entropy. It was found that θ' was the stable phase up to 500 K. Above this temperature, the vibrational contribution drives the free energy of θ below that of θ', and θ becomes the stable intermetallic phase. θ" was metastable in the whole temperature range with a formation energy significantly higher (less negative) than both θ' and θ.

Homogeneous and heterogeneous nucleation of θ" and θ' was evaluated from the saddle point of the free energy surface, which is a function of volume and aspect ratio of the precipitates and depended on the chemical, elastic and interface energy contributions. The chemical free energy was estimated from the CALPHAD database while elastic and interface energies were given by the stress-free transformation strains and first principles calculations, respectively. It was found that the critical free energy and precipitate size for homogeneous nucleation of θ" were much smaller (about two orders of magnitude) than those for θ'. The interaction energy induced by the presence of either a dislocation loop or a θ" precipitate reduced the critical free energy and precipitate size for heterogeneous nucleation in approximately in order of magnitude for θ' precipitates. Nevertheless, this effect was much smaller in the case of θ" precipitates. Thus, precipitation begins with the homogeneous nucleation of θ" precipitates, followed by heterogeneous precipitation of θ' precipitates on dislocations and, afterwards, on pre-existing θ" precipitates. Homogeneous nucleation of θ' precipitates only occurs after longer aging times as well as the progressive transformation of metastable θ" into stable θ' or θ precipitates.

Finally, the growth and equilibrium shape of θ" and θ' precipitates was calculated by means of mesoscale phase field model which included the free energy associated with chemical, interface and elastic strain energy contributions. It was shown that the aspect ratio of the precipitates increased during growth and was higher than the one predicted by Wulff plot. This behaviour was due to the contribution of the elastic strain energy, which is proportional to the precipitate volume, while the interface energy contribution only increases with the area.

The model predictions in terms of precipitate stability, nucleation and orientation and shape of the precipitates were in good agreement with the experimental data in the literature, validating the approach developed in this investigation. Moreover, the analysis of the different phenomena (stability, nucleation, growth) provides a comprehensive picture of the precipitation process in Al-Cu alloys that it is able to rationalize and quantify the experimental observations. Further developments should be aimed at the design of optimum microstructures based on the information obtained from first principles calculations of the different energy contributions.

**Acknowledgements**

This investigation was supported by the European Research Council (ERC) under the European Union's Horizon 2020 research and innovation programme (Advanced Grant VIRMETAL, grant agreement No. 669141). Computer resources and technical assistance provided by the Centro de Supercomputación y Visualización de Madrid (CeSViMa) are gratefully acknowledged. Additionally, the authors thankfully acknowledge the computer resources at Finisterrae and the technical support provided by CeSGa and Barcelona Supercomputing Center (project QCM-2017-3-0007). HL is grateful to the support from the FWO and European Union's 2020 research and innovation programme under the Marie Sklodowska-Curie grant agreement No. 665501.

**Appendix 1**

*Ab initio* calculations were performed using the Quantum Espresso plane-wave pseudopotential code [50]. The fundamental eigenvalues were calculated with the help of the Kohn-Sham approach [51]. The interaction between valence electrons and core electrons was treated under the pseudopotential approximation, and ultrasoft pseudopotentials were used to reduce the basis set of plane wavefunctions used to describe the real electronic functions [52]. The exchange-correlation energy was evaluated with the help of the Perdew-Burke-Erzenhof (PBE) approach, within the Generalised Gradient Approximation (GGA) [53]. After conducting careful convergence tests, it was found that an energy cutoff of 37 Ry (503 eV) was sufficient to reduce the error in the total energy to less than 1 meV/atom. A k-point grid separation of 0.03 Å$^{-1}$ was employed for the integration over the Brillouin zone according to the Monkhorst Pack scheme [54]. A well converged k-point set is required for high quality thermodynamics calculation, thus separate convergence tests were carried out to confirm that the error in the phonon frequencies was less than 10$^{-3}$ cm$^{-1}$. All the phonon eigenfrequencies were found to be real, hence it was confirmed that the compounds were mechanically stable.

The method of finite displacements (also known as the supercell method) [20] was used to determine the DOS of each phase as a function of temperature. This technique which is based on calculating the forces on atoms after perturbing slightly the atomic positions by using single point energy calculations. Each calculation provides *3N* elements of the force constant matrix and symmetries can be used to deduce more elements, reducing the number of calculations needed. The force constant matrix, $\Phi_{i\alpha,j\beta}$ (where *i* and *j* stands for the atoms and *α* and *β* for the Cartesian directions) can be calculated from the forces on atom *j*, $F_{j\beta}$ due to the $u_{i\alpha}$ displacement of atom *i* in direction *α* as,

$$\Phi_{i\alpha,j\beta} = \frac{\partial F_{j\beta}}{\partial u_{i\alpha}} = \frac{\partial^2 E}{\partial u_{i\alpha} \partial u_{j\beta}} \quad (A1)$$

The dynamical matrix $D_{i\alpha,j\beta}(q)$ is obtained by means of the Fourier transform of $\Phi_{i\alpha,j\beta}$ at wavevector *q* from which the eigenfrequencies *ω* are computed. After all eigenfrequencies are computed, the density of normal modes (in a normal mode, all the atoms move with the same frequency) or phonon DOS is calculated. The term $g(\omega)$ is defined as the total number of modes with frequencies between *ω* and *ω* + *dω* per unit volume.

Supercells large enough to contain the sphere for which the force constant matrix includes non-zero elements were employed. That sphere is described by the real space cutoff radius, beyond which it is assumed that no atomic interactions exist. The larger the supercell, the more accurate, but also the more expensive the calculation. Separate convergence tests of the free energy with respect to the cutoff radius were carried out until the error was < 1 meV/atom. The minimum value for the cutoff radius was selected to be 6 Å, which resulted to 3 × 3 × 3 supercells for Cu and Al, 2 × 2 × 3 supercells for θ, and 3 × 3 × 2 supercells for θ' and θ".

**Appendix 2   Chemical driving force for precipitate nucleation**

The chemical driving force for precipitate nucleation can be calculated following the methodology outlined in [25−26]. If a small amount of atoms with the composition of the precipitate phase $x_\mathrm{p}$ ($x_\mathrm{p}$ = 0.25 and 1/3 for θ" and θ', respectively) is removed from the α−Al matrix, the total free energy of the system will decrease by $\Delta G_1^p$:

$$\Delta G_1^p = \mu_{Al}^\alpha(1 - x_p) + \mu_{Cu}^\alpha x_p, \tag{A2}$$

where $\mu_{Al}^\alpha$ and $\mu_{Cu}^\alpha$ stand for the chemical potentials of Al and Cu, respectively, in the α−Al matrix with a Cu content $x_0$. If these atoms are now rearranged to the precipitate structure, the total free energy of the system will increase by $\Delta G_2^p$:

$$\Delta G_2^p = \mu_{Al}^p(1 - x_p) + \mu_{Cu}^p x_p, \tag{A3}$$

where $\mu_{Al}^p$ and $\mu_{Cu}^p$ stand for the chemical potentials of Al and Cu, respectively, in the precipitate with a Cu content $x_p$. Therefore, the nucleation driving force $\Delta G_C^p$ is given by:

$$\Delta G_C^p = \Delta G_2^p - \Delta G_1^p. \tag{A4}$$

From the definition of the chemical potentials [25], it is known that $\mu_{Al}^\alpha$ and $\mu_{Cu}^\alpha$ in eq. (A2) can be expressed as:

$$\mu_{Al}^\alpha = f_\alpha(x_0) - x_0 f_\alpha'(x_0) \text{ and } \mu_{Cu}^\alpha = f_\alpha(x_0) + (1 - x_0) f_\alpha'(x_0) \tag{A5}$$

while $\mu_{Al}^p$ and $\mu_{Cu}^p$ in eq. (A3) are given by:

$$\mu_{Al}^p = f_\alpha(x_{eq}) - x_{eq} f_\alpha'(x_{eq}) \text{ and } \mu_{Cu}^p = f_\alpha(x_{eq}) + (1 - x_{eq}) f_\alpha'(x_{eq}) \tag{A6}$$

After simple algebraic operations, the chemical driving forces for the nucleation of the θ″ and θ′ precipitates, $\Delta G_C^{\theta''}$ and $\Delta G_C^{\theta'}$ in eq. (6), can be derived.

**Appendix 3  Chemical Gibbs free energies of α–Al and θ″**

According to the CALPHAD database [28], eq. (A2) can be used to describe the chemical free energy of α–Al matrix, $f_\alpha$

$$f_\alpha = \sum_{i=Al,Cu} x_i G_i + RT \sum_{i=Al,Cu} x_i \ln x_i + {}^E G. \tag{A7}$$

In this equation, $x_i$ and $G_i$ are the mole concentration and the molar Gibbs free energy of a pure element $i$, respectively. When $i$ = Al and Cu, $G_{Al}$ and $G_{Cu}$ are given by:

$G_{Al}^{fcc} = -7976.15 + 137.093038T - 24.3671976T \ln T - 18.84662 \times 10^{-4} T^2 + 74092 T^{-1}$
$\quad -8.77664 \times 10^{-7} T^3 \quad (298 < T(K) < 700)$

$G_{Cu}^{fcc} = -7770.458 + 130.485235T - 24.112392T \ln T - 26.5684 \times 10^{-4} T^2 52478 T^{-1}$
$\quad -1.29223 \times 10^{-7} T^3 \quad (298 < T(K) < 1358)$

The last term in eq. (A2), ${}^E G$, is the excess Gibbs free energy. It is represented by the Redlich-Kister polynomial:

$$^{E}G = x_{Al}x_{Cu} \sum_{j=0,1,2} {}^{j}L_{AlCu}(x_{Al} - x_{Cu})^{j}, \tag{A8}$$

where ${}^{j}L_{AlCu}$ are given by:

$${}^{0}L_{AlCu} = -53520 + 2T \qquad {}^{1}L_{AlCu} = 38590 - 2T \qquad {}^{2}L_{AlCu} = 1170$$

In this work, a fourth order polynomial is used to fit $f_\alpha$ at $T = 200$ °C in a dimensionless form:

$$f_\alpha(x) = -1.4632 - 2.9571x - 3.9656x^2 + 5.8588x^3 + 0.8350x^4, \tag{A9}$$

The chemical free energy function of the metastable phase θ", $f_{\theta''}$, was approximated by a parabolic function of the Cu concentration since $f_{\theta''}$ is not available from the CALPHAD database. The $f_{\theta''}$ and $f_\alpha$ curves have a common tangent line, which goes from the $f_\alpha$ curve at the equilibrium concentration of Cu in the α–Al matrix ($x_0 = 0.001$) to that of θ" ($x_{eq} = 0.25$). The only degree of freedom of $f_{\theta''}$ is its curvature and the chemical free energy of θ" is given by

$$f_{\theta''}(x) = 0.4182 - 1.9510x + x^2. \tag{A10}$$

## Appendix 4  Relationship between dimensionless parameters used in phase field simulations and physical parameters

The dimensionless parameters used in phase-field model and the corresponding physical parameters are linked by the grid size through the scaling between the dimensional and dimensionless elastic and interfacial energies [55−57]. The dimensionless and physical parameters of the phase field model for θ" are given in Table A1. The parameters used to simulate the evolution of θ' are given in [30].

Table A1. Dimensionless parameters for the phase-field model and the corresponding physical parameters.

| Parameter | Symbol | Dimensionless | Physical |
|---|---|---|---|
| Elastic constants | $C_{11}$ | 110.4 | 110.4 GPa |
|  | $C_{12}$ | 60 | 60.0 GPa |
|  | $C_{44}$ | 31.6 | 31.6 GPa |
| Molar Volume | $V_m$ | 1 | $10^{-5}$ m$^3$mol$^{-1}$ |
| Grid size | $l_0$ | 1 | 0.5 nm |
| Interfacial energy | $\Gamma$ | 1 | 166.67 J/m$^2$ |
| Interface mobility | $M$ | 5 | - |
| Order parameter mobility | $L$ | 1 | - |
| Domain wall coefficients | $B$ | 10 | - |

The dimensionless and physical parameters for chemical free energy are shown in Appendix 3.

In addition, the computational time step $t^*$ and the real time $t$ are related by the dimensional and dimensionless interfacial mobility $M$ and $M^*$, the grid size, $l_0$, and the characteristic free energy $\Delta f$ (usually the maximum driving force for the phase transformation from the bulk free energy) according to [32]:

$$t = \frac{t^* M^* l_0^2}{M|\Delta f|}, \quad (A11)$$

where $M$ can be expressed as an averaged mobility $D\bar{c}(1-\bar{c})$, where $D$ is the diffusion coefficient, and $\bar{c}$ is the average Cu concentration in α matrix, which is a function of evaluation time. Nevertheless, the comparison of the simulation time with the real aging time was out of the scope of the paper.

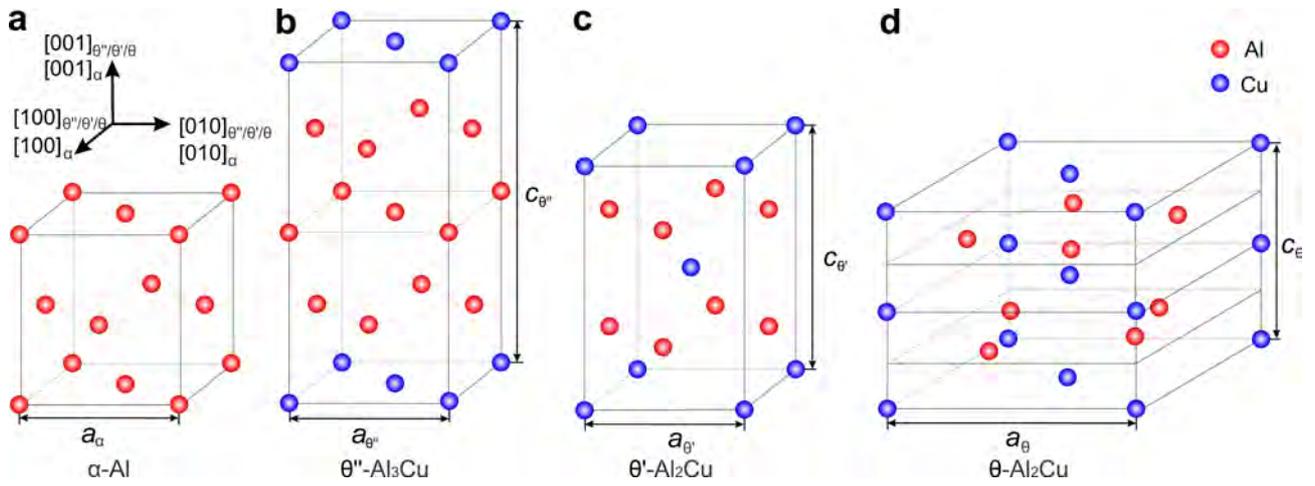

Fig. 1. Unit cells of α–Al matrix and θ'', θ' and θ intermetallic phases.

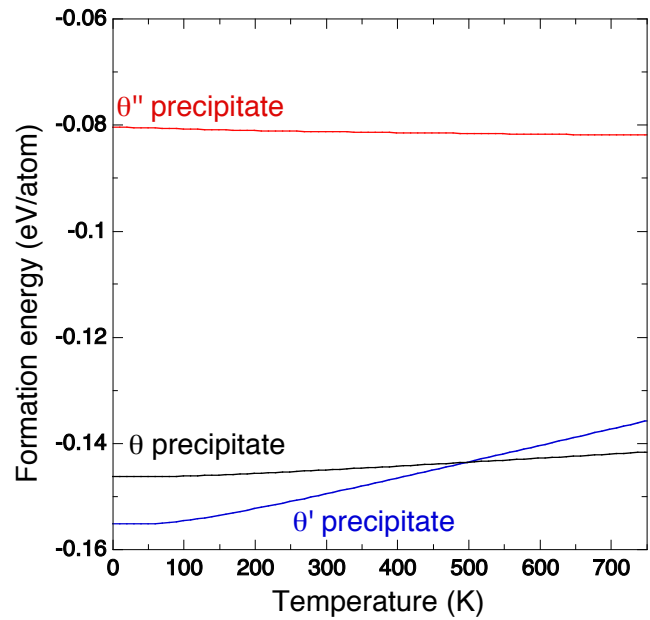

Fig. 2. Formation energy of θ'', θ' and θ intermetallic phases as a function of temperature obtained from first principles calculations.

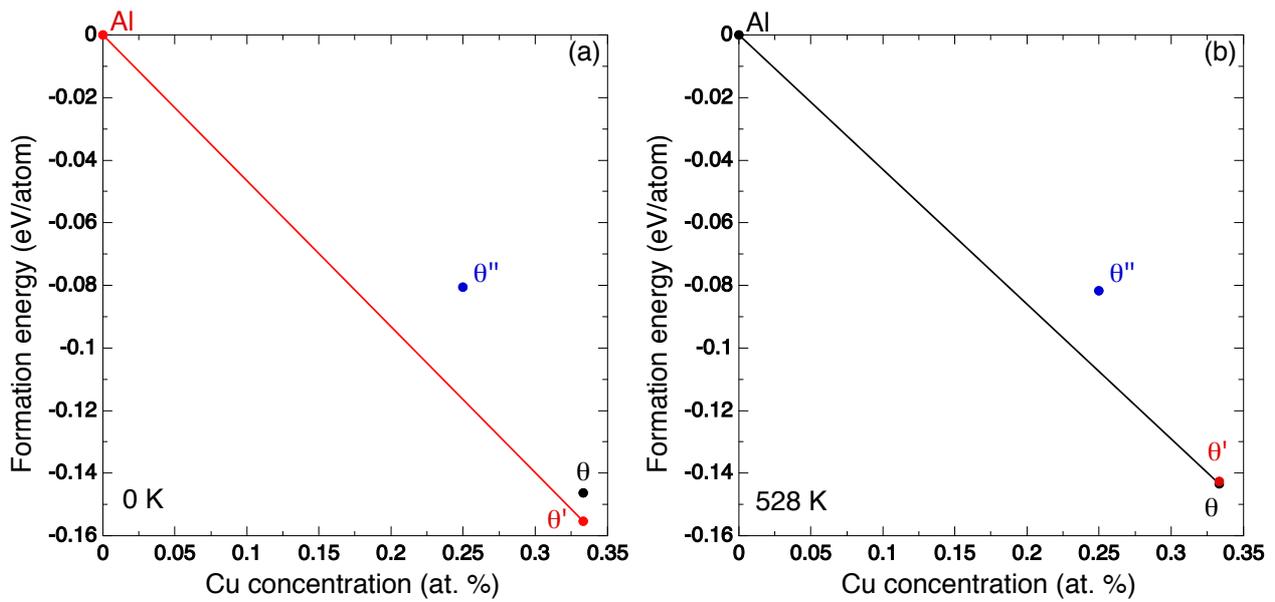

Fig. 3 Convex hull analysis of the thermal stability of θ″ (Al$_3$Cu), θ′ (Al$_2$Cu) and θ (Al$_2$Cu) phases at different temperatures. (a) 0 K. (b) 528 K.

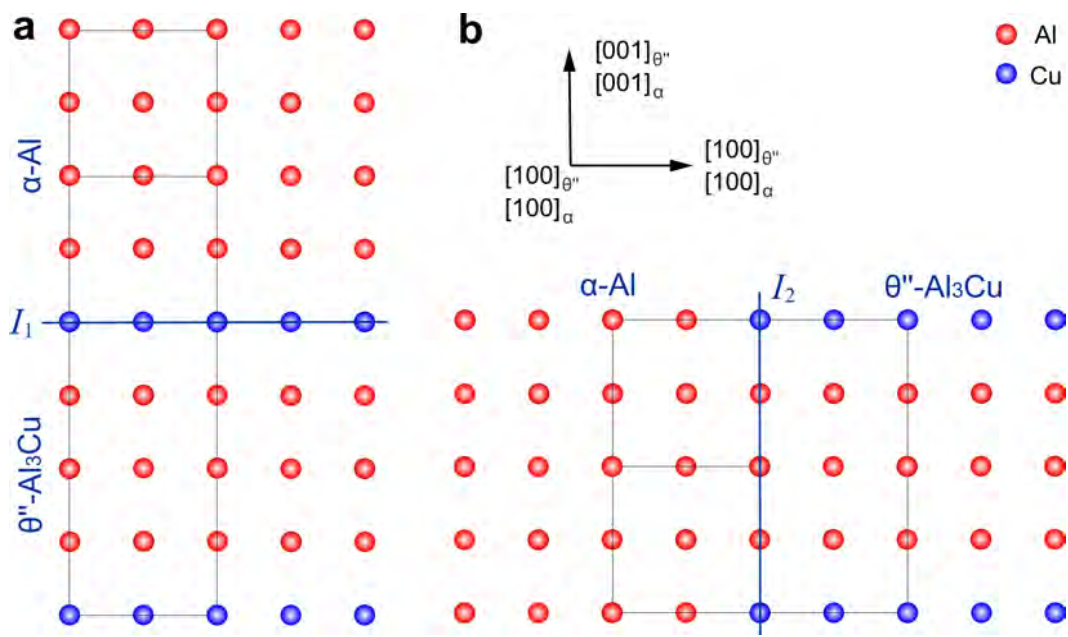

Fig. 4. Coherent interfaces between θ″ (Al$_3$Cu) and α−Al matrix. (a) Type $I_1$. (b) Type $I_2$.

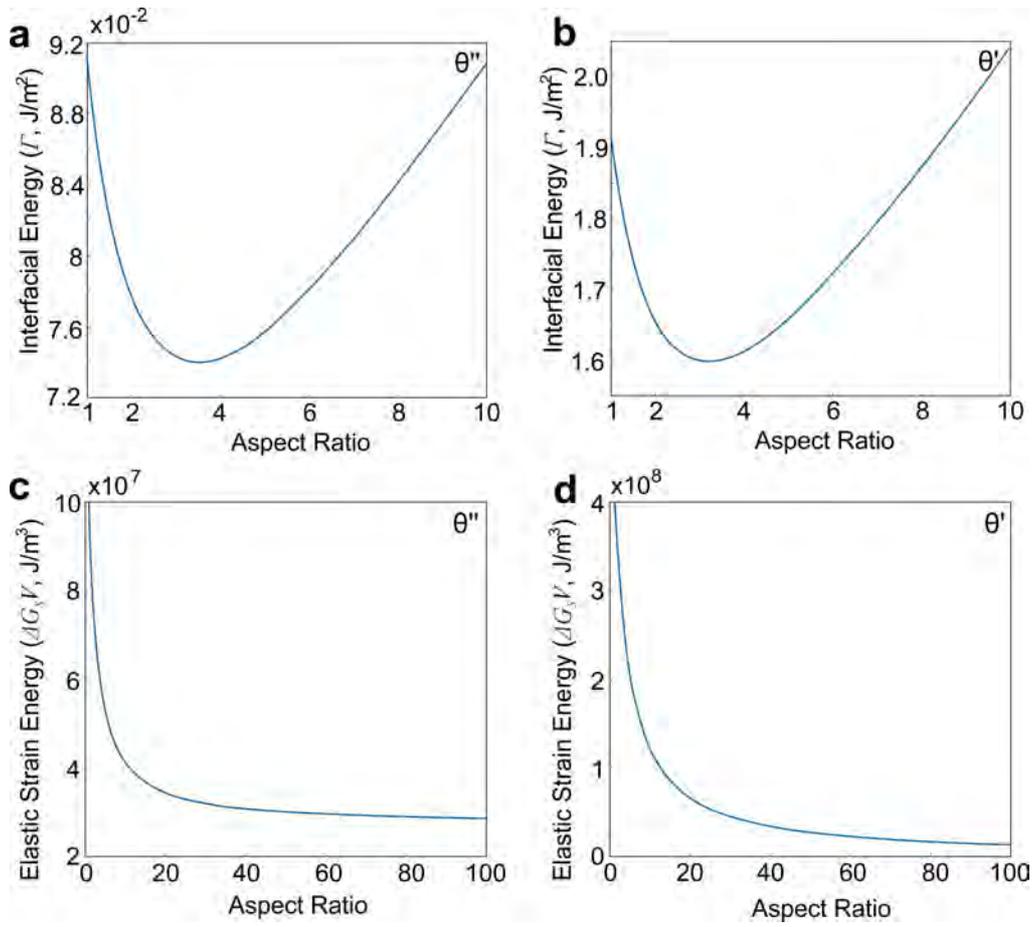

Fig. 5. Influence of the aspect ratio $A=m_1/m_0$ on the interfacial and elastic strain energies for the nucleation of precipitates of unit volume $V = 1$. (a) and (c) $\theta''$ precipitate. (b) and (d) $\theta'$ precipitate.

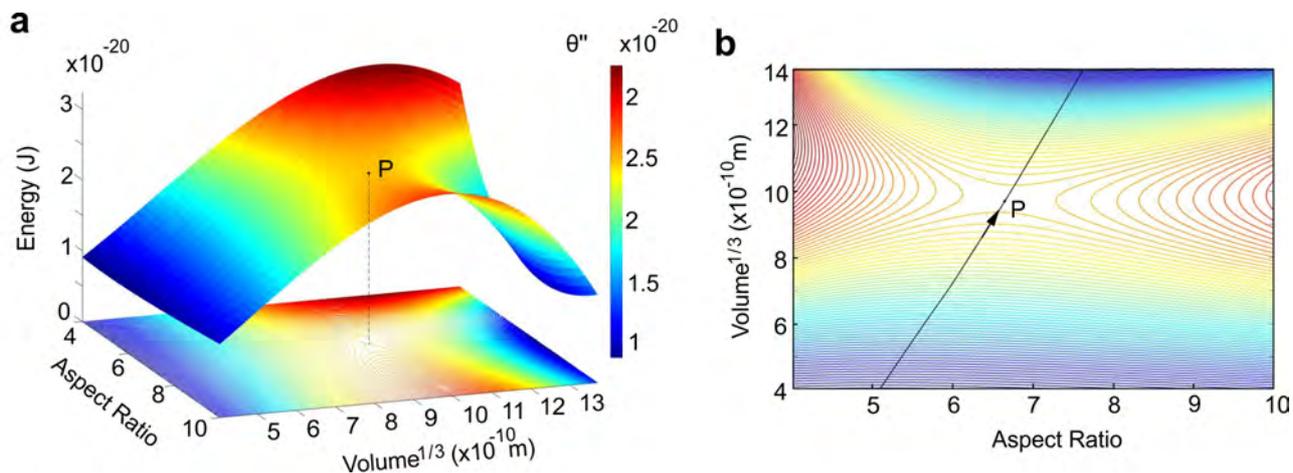

Fig. 6 $\Delta G$ surface for the homogeneous nucleation of $\theta''$ precipitates including the contribution of chemical, interfacial and elastic strain energies. The saddle point of $\Delta G$ surface is marked by $P$ in each figure. The 2D projection of the $\Delta G$ surface (a) is shown in (b). The black line with an arrow in (b) shows the minimum energy path for the nucleation of $\theta''$ precipitates.

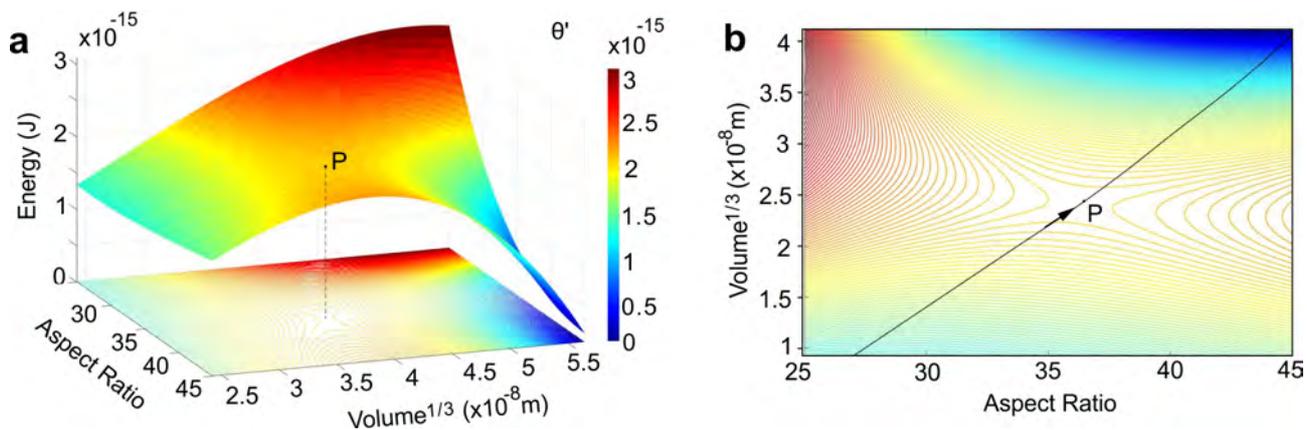

Fig. 7 $\Delta G$ surface for the homogeneous nucleation of $\theta'$ precipitates including the contribution of chemical, interfacial and elastic strain energies. The saddle point of $\Delta G$ surface is marked by $P$ in each figure. The 2D projection of the $\Delta G$ surface (a) is shown in (b). The black line with an arrow in (b) shows the minimum energy path for the nucleation of $\theta''$ precipitates.

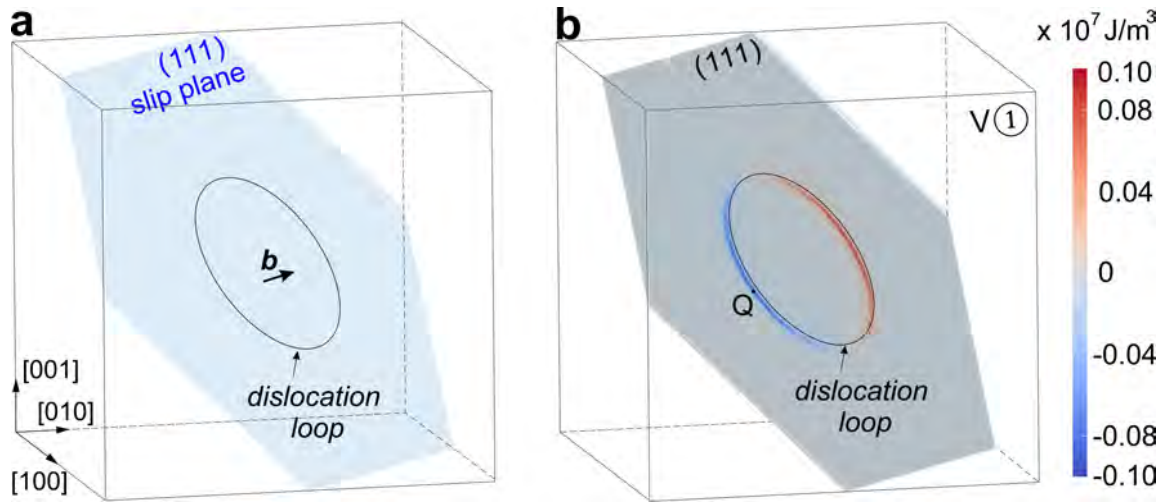

Fig. 8 (a) A dislocation loop with a $a_\alpha\langle\bar{1}10\rangle/2$ Burgers vector in a $(111)_\alpha$ slip plane. (b) The distribution of interaction energy between the stress field of the dislocation loop and the SFTS of θ" precipitates belonging to variant ① in a $(111)_\alpha$ plane at 0.25 nm below the slip plane. The minimum (most negative) interaction energy is marked by Q.

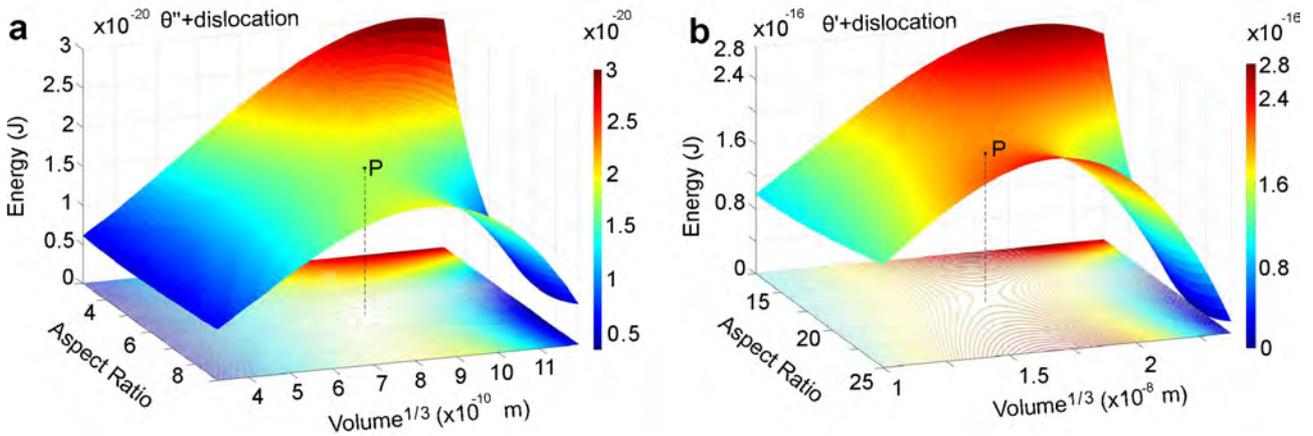

Fig. 9 $\Delta G$ surfaces for the heterogeneous nucleation of (a) θ" and (b) θ' precipitates on a dislocation loop. The saddle point of the $\Delta G$ surface is marked by $P$ in each figure.

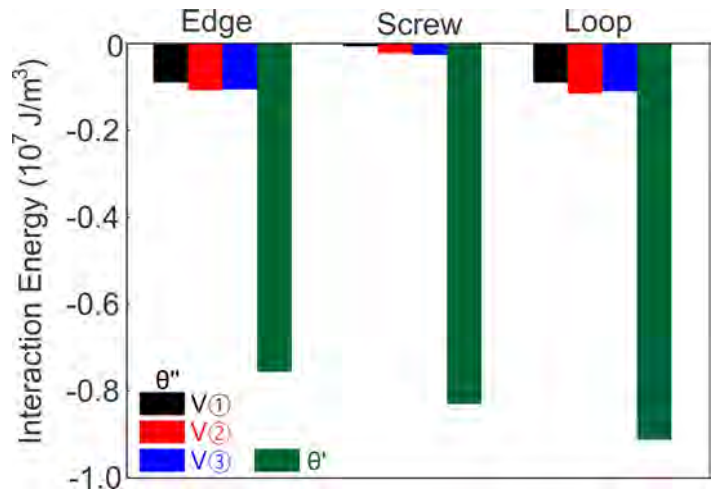

Fig. 10 Minimum interaction energies between the stress field of edge and screw dislocations and a dislocation loop $\boldsymbol{b} = a_0[\bar{1}10]_\alpha/2$ in the $(111)_\alpha$ plane and three orientation variants of θ″ precipitates. These interaction energies are compared with the minimum interaction energies when θ′ form on these dislocations.

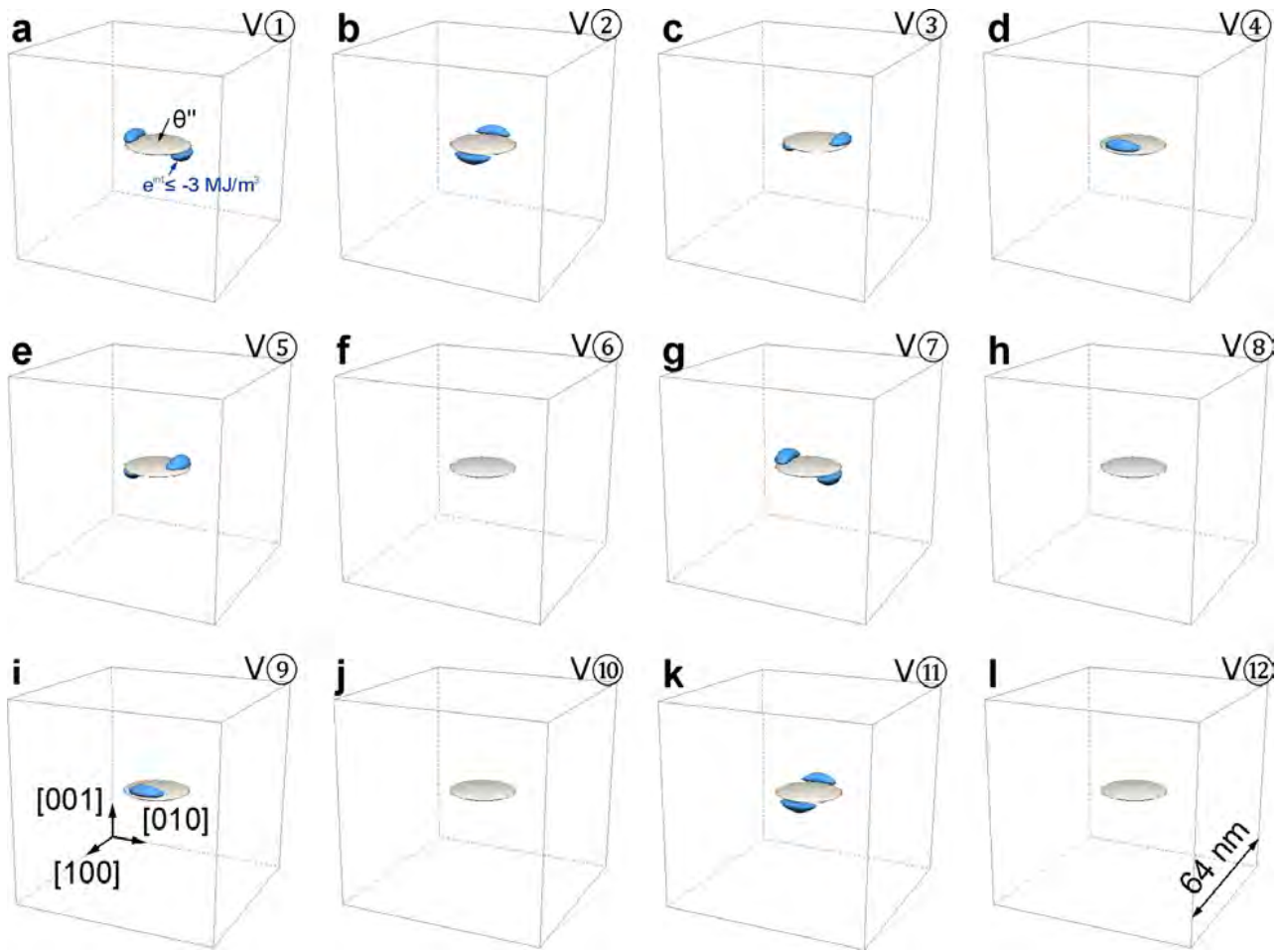

Fig. 11 Interaction energy between the stress field of the pre-existing θ″ precipitate belonging to variant ① and the SFTS of different variants of the θ′ precipitate. The blue zones show the regions around the θ″ precipitate where the interaction energies are more negative than −3 MJ/m³.

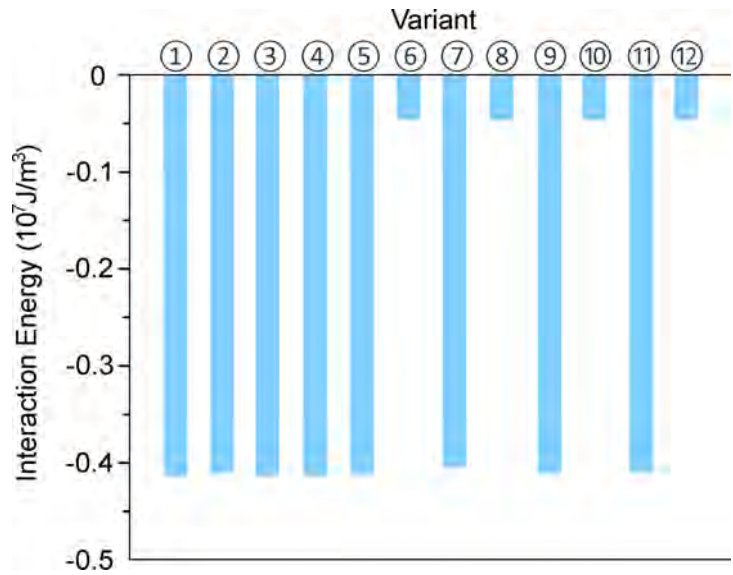

Fig. 12 Minimum interaction energies between the stress field of the pre-existing θ" precipitates belonging to variant ① and the SFTS of the to-be-nucleated θ' precipitate.

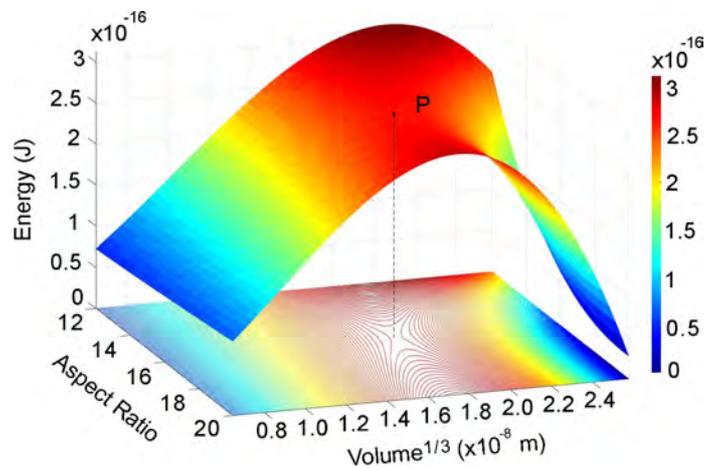

Fig. 13 $\Delta G$ surfaces for the heterogeneous nucleation of θ' precipitates on a θ" precipitate. The saddle point of the $\Delta G$ surface is marked by $P$.

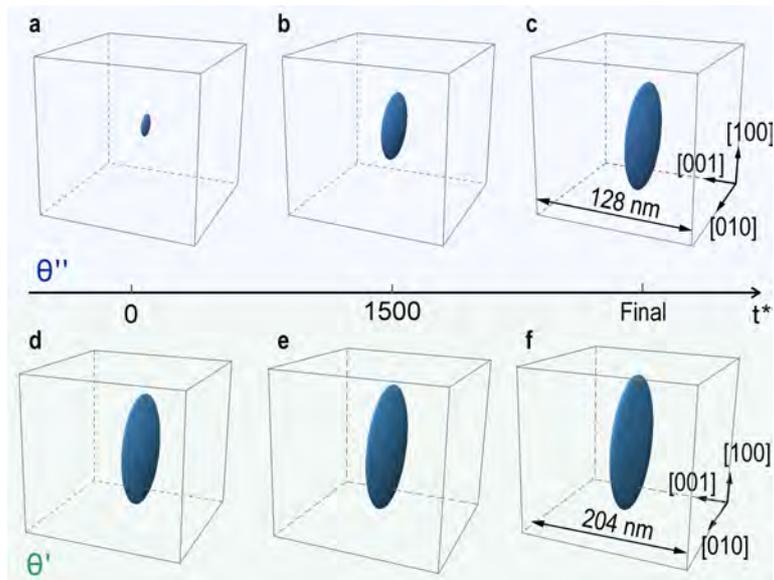

Fig. 14 Simulation results of the evolution of (a-c) a θ" precipitate and (d-f) a θ' precipitate in Al-1.74 at.% Cu alloy aged at 190 °C.

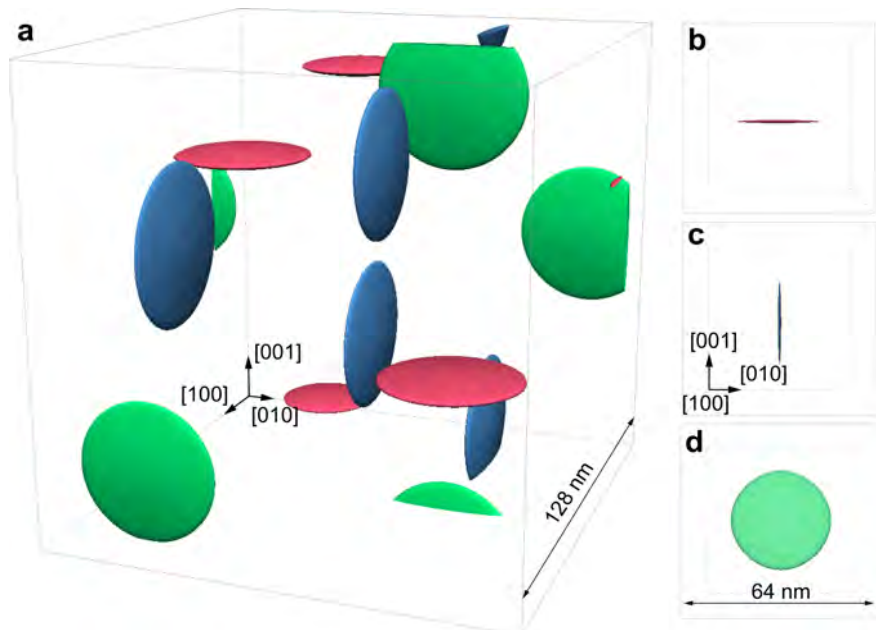

Fig. 15 (a) The distribution of homogeneously formed θ" precipitates and the shape of variants (b) ①, (c) ② and (d) ③.

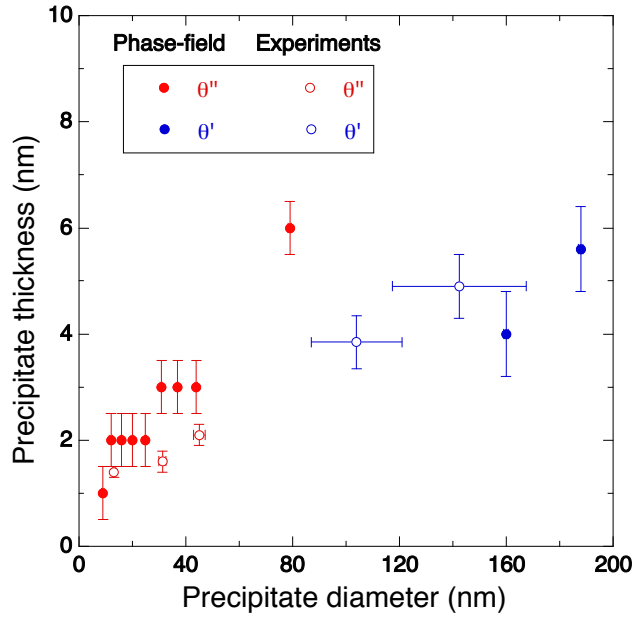

Fig. 16 Comparison between experimental results and phase-field simulations of the shape of θ″ and θ″ precipitates (as given by the average diameter and thickness). The experimental data were obtained in an Al–1.74 at.% Cu alloy [12, 29].

Table 1 Interfacial energies (in mJ/m$^2$) of θ″/ α-Al, θ′/α-Al obtained from density functional theory simulations.

| Interface | Type | Energy (mJ/m$^2$) |
|---|---|---|
| $\theta'$/ α-Al | coherent | 152 |
| $\theta'$/ α-Al | semi-coherent | 487 |
| $\theta''$/ α-Al | $I_1$ | 6.6 |
| $\theta''$/ α-Al | $I_2$ | 23.3 |

Table 2. Nucleation energy barrier, critical aspect ratio, length, thickness and volume of the nucleus of θ″, θ′, and θ′ formed by homogeneous nucleation or heterogeneous nucleation in a dislocation loop.

| Nucleation | Energy Barrier (J) | Aspect Ratio | Thickness (nm) | Length (nm) | Volume (nm$^3$) |
|---|---|---|---|---|---|
| Homogeneous θ″ | 2.5 10$^{-20}$ | 6.7 | 0.34 | 2.3 | 0.95 |
| Homogeneous θ′ | 2.1 10$^{-15}$ | 36.2 | 4.4 | 160 | 59365 |
| Heterogeneous θ″ | 1.8 10$^{-20}$ | 6.0 | 0.31 | 1.9 | 0.56 |
| Heterogeneous θ′[a] | 2.5 10$^{-16}$ | 21.0 | 2.7 | 55.7 | 4314 |
| Heterogeneous θ′[b] | 7.0 10$^{-16}$ | 29.0 | 3.2 | 92.2 | 14137 |

[a] θ′ forms on dislocation loops
[b] θ′ form on θ″ precipitates

## Supplementary Materials

### Explicit nucleation

A nucleus forming in an individual grid cell is modelled as a simple Bernoulli trial. Each cell contains a number of atomic sites, and each one of them could be a nucleation site. The probability of forming a nucleus at one atomic site in one characteristic nucleation time interval can be calculated, according to the classical theory [1-3], as

$$J^* = ZN\beta^* e^{-\frac{\Delta G^*}{kT}} e^{\frac{\tau}{t}} \qquad (S1)$$

where $J^*$ is the nucleation rate per cell, $Z$ the non-equilibrium factor due to Zeldovich, $N$ the number of atoms in each grid cell, $\beta^*$ a frequency factor equal to the reciprocal of the characteristic nucleation time, $\Delta G^*$ the nucleation energy barrier of the precipitate nucleus, $k$ Boltzmann's constant, $T$ the ageing temperature, $\tau$ the incubation time and $t$ the time. By neglecting the incubation time [3], eq. (S1) is simplified to

$$J^* = \kappa_1 e^{-\frac{\Delta G^*}{kT}}, \qquad (S2)$$

where $\kappa_1 = ZN\beta^*$. In this work, $T = 453$ K, $\kappa_1$ was assumed to be $10^{-6}$, and $\Delta G^*$ for θ" and θ' can be found in Table 2, leading to a nucleation rate of approximately $10^{-8}$ per grid cell in the characteristic nucleation time. More detailed information is given in [3].

The probability of forming one nucleus per atomic site during the $1/\beta^*$ time, $p$, is expressed as [4−6]:

$$p = \frac{J^*}{N\beta^*}. \qquad (S3)$$

In this investigation, the nucleation process was modelled with a random number generator. This random number (in the range 0 to 1) was generated for each cell in each time step. If the random number was lower than $p$, it was assumed the formation of a nucleus in the grid and otherwise no nucleus was created.